\documentstyle[multicol,prd,aps,graphicx]{revtex} 
\begin{document} 
\draft

\title{  
%\vglue -0.5cm 
%\hfill{\small IFT-P.032/2003} \\ 
%\hfill{\small hep-ph/0308037} \\ 
%\hfill{\small August 2003}\\ 
%\vglue 0.5cm 
Stabilizing the invisible axion in 3-3-1 models     
}  
 
\author{Alex G. Dias$^1$ and V. Pleitez$^2$} 
\address{$^1$Instituto de F\'\i sica, Universidade de S\~ao Paulo, \\ 
05315-970  S\~ao Paulo, SP, Brazil}  
\address{$^2$Instituto de F\'\i sica Te\'orica, Universidade Estadual 
Paulista,\\  
Rua Pamplona 145, \\  
01405-900 S\~ao Paulo, SP, Brazil}  
 
\date{\today} 
\maketitle

\begin{abstract} 
By introducing local $Z_N$ symmetries with $N=11,13$  in 
two 3-3-1 models, it is possible to implement an automatic Peccei-Quinn 
symmetry, keeping the axion protected against gravitational 
effects at the same time. Both models have a $Z_2$ domain wall problem and 
the neutrinos are strictly Dirac particles.  
 
\end{abstract} 
\pacs{PACS numbers: 14.80.Mz; 12.60.Fr; 11.30.Er } 
%14.80.Mz Axions and other Nambu-Goldstone bosons  
%(Majorons, familons, etc.)     
% 
%12.60.Fr Extensions of electroweak Higgs sector 
%11.30.Er Charge conjugation, parity, time reversal,  
%and other discrete 
%symmetries 
%23.40.-s beta decay; double beta decay; electron and muon capture 
%12.60.Cn Extensions of electroweak gauge sector 
%14.60.St Non-standard-model neutrinos, right-handed neutrinos, etc. 

\begin{multicols}{2} 
\narrowtext 
\section{Introduction} 
\label{sec:intro} 
 
Recently, observations of the core mass distribution in the cluster of Galaxies 
Abell 2029 using the NASA's Chandra X-ray Observatory suggest the
existence of 
cold dark matter (CDM)~\cite{cdm}. On the other hand, the Wilkinson Microwave 
Anisotropy Probe (WMAP) measurements of the cosmic microwave background 
temperature anisotropy and polarization are also consistent with CDM and a 
positive cosmological constant~\cite{wmap}. 
Although the exact nature of the CDM is not known yet, candidates for this sort 
of matter are elementary particles such as neutralinos or the 
invisible axion~\cite{darkmatter}.  
However, early 
invisible axion models~\cite{dine,kim} are unstable against quantum 
gravitational effects~\cite{gravity}, which may generate a large axion mass and 
also spoil the value of the $\bar{\theta}_{\rm eff}$ parameter. One way to 
stabilize the axion is by considering large discrete gauge symmetries in the 
sense of Ref.~\cite{kw} as  was done in the multi-Higgs extension of the standard
model~\cite{axionsm}, in 
the 3-3-1 model~\cite{axion331} or in the supersymmetric model~\cite{babu}.    
The search for dark matter is of course related to the search 
for new physics beyond the standard model which in turn  is related to the 
e\-xis\-ten\-ce of new fundamental energy scales. In the literature, the most easily 
recognized fundamental energy scales are those related to supersymmetry, 
the neutrino masses, grand unification, and superstring theory.  
 
In this vein, it is worth  recalling once more that it has been known for a 
long time that the measured value of the electroweak mixing angle 
$\sin^2\theta_W(M_Z)=0.23113\lesssim1/4$ appears to obey, at an energy scale 
$\mu$, an $SU(3)$ symmetry in such a way that 
$\sin^2\theta_W(\mu)=1/4$~\cite{sw72}. Hence, if the value of $\sin^2\theta_W$  
is not an accident, it may be considered as an indication of a 
new fundamental energy scale of the order of a few TeVs. Notwithstanding, in 
models with $SU(3)$ electroweak symmetry there is trouble when we try to 
incorporate quarks. A solution to this issue is to introduce an extra $U(1)$ 
factor such as in 3-3-1 models~\cite{331,outros}, to embed the model in a 
Pati-Salam-like model~\cite{dimo}, or even to embed it in theories of TeV 
gravity~\cite{extra}.    
 
Independently of the axion or dark matter issues, 3-3-1 models are interesting 
possibilities, on their own, for  physics at the TeV scale.  At low energies 
they coincide with the standard model and some of them give at a least 
partial explanation of some fundamental questions that are accommodated but not 
explained by the standard model. For instance, {\it i)} in order to cancel the 
triangle anomalies the number of generations  
must be three or a multiple of three; {\it ii)} the model of Ref.~\cite{331} 
predicts that $(g'/g)^2=\sin^2\theta_W/(1-4\sin^2\theta_W)$; thus there is a 
Landau pole at the energy scale $\mu$ at which $\sin^2\theta_W(\mu)=1/4$, 
and according to recent calculations $\mu\sim4$ TeV~\cite{phf}; {\it iii)} the  
quantization of the electric charge~\cite{pr} and the vectorial
cha\-rac\-ter of the electromagnetic interactions~\cite{cp} do not depend
on the nature of the neutrinos, i.e., whether they are Dirac or Majorana particles;  
and {\it iv)} the model possesses ${\cal N}=1$ supersymmetry naturally 
at the $\mu$ scale~\cite{331s}. If right-handed neutrinos are considered 
to transform nontrivially, 3-3-1 models~\cite{331,outros} can be embedded 
in a model with 3-4-1 gauge symmetry in which leptons 
transform as $(\nu_l,l,\nu^c_l,\,l^c)_L\sim({\bf1},{\bf4},0)$ under each gauge 
factor~\cite{su4}.      
 
Models with  $SU(3)$ (or $SU(4)$) symmetry may have doubly charged 
vector bosons. These types of bileptons may be detected by 
measuring the left-right asymmetries in M\o ller scattering~\cite{assi}, for 
instance, at the E158 SLAC experiments (which use 48 GeV polarized  
electrons scattering off unpolarized electrons in a liquid hydrogen 
target~\cite{e158}); or in future lepton-lepton accelerators.  
It is interesting that the weak interaction's parity nonconservation has never 
been observed in lepton-lepton scattering.  
Those asymmetries may also be used for seeking a heavy neutral 
$Z^{\prime 0}$ vector boson, which is also a prediction of these models, in 
$e\mu$ collisions~\cite{emu}. Singly and doubly charged vector bileptons
may also be produced in $e^-\gamma$~\cite{egamma} or
$\gamma\gamma$~\cite{gammagamma} or hadron ~\cite{dion} colliders.  
New heavy quarks are also part of the electroweak quark multiplets in the
minimal model representation. They are singlets under the  standard model
$SU(2)_L\otimes U(1)_Y$ group 
symmetry. In some versions their electric charge is  different from the
usual one, so that it can be used to distinguish such a model from
their viable competitors. In fact, 
the $p\overline{p}$ production and decay of these exotic quarks at the energies 
of the Tevatron have been studied in Ref.~\cite{djm} where a lower bound
of 250 GeV on their masses was found.
This sort of models is also predictive with respect to neutrino 
masses~\cite{numass}; the models can implement the large mixing angle 
MSW solution to the solar neutrino issue~\cite{lma}, and also the almost 
bimaximal mixing matrix in the lepton sector~\cite{pmns}. 

Summarizing, from the present experimental data, say those from the CERN $e^{+}e^{-}$ 
collider LEP, 3-3-1
models are safe if the symmetry breaking from 3-3-1 to 3-2-1 occurs at
the level of TeVs; however, they 
have  rich phenomenological consequences as we mentioned above. It
will be interesting to search for some of the new particles that are
present in these models, 
as extra Higgs scalars, exotic quarks and vector bileptons, at the
energies of the upgrade DESY $ep$ collider HERA and Tevatron~\cite{dion,mk}. 
The scalar sectors are equivalent to multi-Higgs-boson 
extensions of the standard model; for instance, under $SU(2)_L\otimes
U(1)_Y$ the model with three triplets has two doublets and several
non-Hermitian singlets, while the model with a sextet has three doublets, a
complex triplet, and several complex singlets. In particular 
the  neutral singlet ($\chi^0$) is $Z$-phobic (its coupling with $Z^0$
vanishes when the scale of the $SU(3)_L$ symmetry goes to infinity) and
for this reason it evades the LEP constraints. For a finite $SU(3)_L$ energy
scale there are corrections that can be calculated by using the oblique S,
T, and U radiative parameters which constrain the allowed masses for the
leptoquarks and bileptons~\cite{stu}. These masses are
of the same order of magnitude, a few TeV, as those allowed by 
the running of the coupling constants. Through the condition
$\sin^2\theta_W(\mu)=0.25$, the running is sensitive to a new degree of freedom.
Hence, the masses of exotic scalars and bileptons run from hundreds of GeV to a few
TeV~\cite{running}. We will return to this point later.     
 
Turning back to the axion, the interesting point is that a Peccei-Quinn (PQ) 
symmetry~\cite{pq} is almost automatic in the classical Lagrangian of 3-3-1 
models. It is only necessary to avoid a trilinear term in the scalar 
potential by introducing a $Z_2$ symmetry~\cite{pal}. Unfortunately, even in 
this case the PQ symmetry is explicitly broken by gravity effects.    
In order to stabilize the axion, and at the same time 
automatically implement the PQ symmetry, we must introduce local discrete
symmetries, $Z_N$. In fact, recently it was shown that in a version of
the Tonasse and Pleitez 3-3-1 model~\cite{outros} it is possible to
implement  both symmetries $Z_{13}$ and PQ automatically, thus the axion
is naturally light and there is no domain wall problem~\cite{axion331}.   
 
We will consider in this work two 3-3-1 models in which only the known leptons 
transform nontrivially under the gauge symmetry, as in Refs.~\cite{331}, but we  
add also right-handed neutrinos and exotic charged leptons transforming as 
singlets. In one model (model A) we consider a scalar sextet but it is
possible to use only three scalar triplets (model B). Both models admit a
large enough discrete $Z_N$ symmetry, implying a natural light invisible
axion.   
 
\section{The axion in two 3-3-1 models} 
\label{sec:331m} 
 
We will consider two versions of the 3-3-1 model of Ref.~\cite{331}. In model A 
we use three scalar triplets and a sextet, while in model B we avoid the scalar 
sextet. In both models we introduce also a scalar singlet, 
$\phi\sim({\bf1},{\bf1},0)$, and lepton singlets.   
 
The quark and lepton sectors have the same representation content in both models.  
We have quarks transforming, under $SU(3)_C\otimes SU(3)_L\otimes U(1)_N$, as
follows:
$Q_{mL}=(d_m,\, u_m,\, j_m)^T_L\sim ( {\bf3},  
{\bf3}^{*},- 1/3),\;  
m=1,2$; $Q_{3L}=(u_3,\, d_3,\,J)^T_L\sim ( {\bf3}, {\bf 3}, 2/3)$, 
and the corresponding right-handed components in singlets, 
$u_{\alpha R}\sim({\bf3},{\bf1},2/3)$, $d_{\alpha R}\sim({\bf3},{\bf1},-1/3)$, 
$\alpha=1,2,3$; $J_{R}\sim({\bf3},{\bf1},5/3)$; 
$j_{mR}\sim({\bf3},{\bf1},-4/3)$; the leptons are the known ones and transform 
as triplets $({\bf 3}_a,0)$, 
$\Psi_{aL}=(\nu_a,\,l_a,\,l^c_a)^T_L$; $a=e,\mu,\tau$, 
and we also add right-handed neutrinos  and a charged lepton in the 
singlets $\nu_{aR}\sim({\bf1},{\bf1},0)$, $E_{L,R}\sim 
({\bf1},{\bf1},-1)$~\cite{duong,seesaw}. The scalar sector, in the minimal 
version, has only three triplets $\chi=(\chi^{-},\,\chi ^{--},\, 
\chi^0)^T$, $\rho =(\rho^{+},\,\rho ^0,\,\rho ^{++})^T$, $\eta =(\eta^0,\,\eta 
_1^{-},\, \eta _2^{+})^T$, transforming as $({\bf1}, {\bf 3},-1),( {\bf1}, {\bf 
3},1)$ and  $({\bf1},{\bf3},0)$, respectively, and a scalar singlet 
$\phi\sim({\bf1},{\bf1},0)$.  
 
With the quark and scalar multiplets above we have the Yukawa 
interactions 
\begin{eqnarray} 
-{\cal L}^q_Y&=&  
\overline{Q}_{iL} ( F_{i\alpha }u_{\alpha 
R}\rho ^{*}+\widetilde{F}_{i\alpha }d_{\alpha R}\eta ^{*})  
+ \lambda _{im}\overline{Q}_{iL}j_{mR}\chi ^{*}\nonumber \\ 
&+& \overline{Q}_{3L} ( G_{1\alpha } 
u_{\alpha R}\eta + 
\widetilde{G}_{1\alpha }d_{\alpha R}\rho) +  
\lambda _1\overline{Q}_{3L}J_{1R}^{\prime }\chi\nonumber \\ 
&+&H.c., 
\label{yu1} 
\end{eqnarray} 
where repeated indices mean summation.  
 
\subsection{Model with a scalar sextet (Model A)} 
\label{subsec:modela} 
 
In this model we add a scalar sextet $S\sim({\bf1},{\bf6},0)$ with the
following electric charge assignment: 
\begin{equation} 
S=\left(\begin{array}{lll} 
\sigma^0_1 & h^-_1 & h^+_2\\ 
h^-_1 & H^{--}_1 & \sigma^0_2 \\ 
h^+_2 & \sigma^0_2 & H^{++}_2  
\end{array}\right), 
\label{sextet} 
\end{equation} 
and we will assume that only $\sigma^0_2$ gets a nonzero vacuum expectation 
value (VEV) in order to give the correct mass  to the known charged leptons 
plus a mixing with the heavy leptons ($K_a$ and $K^\prime_a$ terms below). 
The Yukawa interactions in the lepton sector are given by   
\begin{eqnarray} 
-{\cal L}^l_Y&=&H^\nu_{ab}\overline{\Psi}_{aL} 
\nu_{bR}\,\eta +H^l_{ab} 
\overline{\Psi}_{aL}S(\Psi_{bL})^c 
+K_a\overline{\Psi}_{aL}E_R\rho \nonumber \\ &+& 
K^\prime_a \chi^T \;\overline{E}_L\,(\Psi_{aL})^c+G_E\overline{E}_LE_R\,\phi 
+H.c.  
\label{yu2} %yu3 
\end{eqnarray} 
where $H^l_{ab}$ is a symmetric matrix in the generation space; we have omitted 
$SU(3)$ indices. Neutrinos are strictly Dirac particles since the total
lepton number will also be an automatically conserved.   
 
Next we impose a $Z_{13}$ discrete symmetry under which the fields 
transform as $Q_{iL}\to \omega^{-1}_2Q_{iL}$, $Q_{3L}\to\omega_0Q_{3L}$, 
$u_{\alpha R} \to \omega_3u_{\alpha R}$, 
$d_{\alpha R}\to \omega^{-1}_5 d_{\alpha R}$, 
$J_R\to \omega_4 J_R$, $j_{mR}\to\omega^{-1}_6 j_{mR}$,  
$\Psi_L\to \omega_6\Psi_L$, $E_L\to\omega_3E_L$,   
$\nu_R\to\omega^{-1}_4\nu_R$, $E_R\to\omega_1E_R$, 
$\eta\to \omega^{-1}_3\eta$, $\rho\to \omega_5\rho$, 
$\chi\to \omega^{-1}_4\chi$,   
$S\to \omega^{-1}_1S$, $\phi\to\omega_2\phi$, where $\omega_k=e^{2\pi 
ik/13},\,k=0,\cdots,6$. Notice that if $N$ is a prime number the singlet $\phi$ 
can transform under this symmetry with any assignment (but the trivial one), 
o\-ther\-wi\-se we have to be careful with the way we choose the singlet $\phi$ 
to transform under the $Z_N$ symmetry. This symmetry implies that the
lowest order effective operator that contributes to the axion mass is
$\phi^{13}/M^9_{\rm Pl}$ which gives a mass of the order 
$(v_\phi)^{11}/M^9_{\rm Pl}$ and also keeps the $\bar{\theta}$ parameter
small as discussed in Ref.~\cite{axion331}. 
 
The most general scalar potential invariant under the gauge and $Z_{13}$ 
discrete symmetries is 
\begin{equation} 
V^{(A)}_{\rm 331} =  V_{\rm H} + 
\left(\lambda_{\phi 1}\,\phi\,\epsilon^{ijk}\eta_i\rho_j\chi_k +  
\lambda_{\phi 2} \chi^TS^\dagger\rho\phi^*+ 
\mbox{H. c.}\right),  
\label{pea} 
\end{equation} 
where $V_{\rm H}$ denotes the Hermitian terms of the potential. This scalar 
potential has the correct number of Goldstone bosons and an axion field. 
 
After imposing the $Z_{13}$ symmetry defined above we have that 
both  the total lepton number $L$ and the PQ symmetry are automatic. The PQ 
charge assignment is as follows: 
\begin{eqnarray} 
u'_L= e^{-i\alpha X_u}u_L,\; d'_L= e^{-i\alpha X_d}d_L,\;\; 
l'_L= e^{-i\alpha X_l}l_L,\nonumber \\ 
\nu'_L=e^{-i\alpha X_\nu}\nu_L,\; 
j'_L=e^{-i\alpha X_j}j_L,\; 
J'_L= e^{-i\alpha X_J}J_L,\; \nonumber \\ 
E'_L=e^{-i\alpha X_{E_L}}E_L,\; 
E'_R=e^{-i\alpha X_{E_R}}E_R, 
\label{pq1} 
\end{eqnarray} 
and in the scalar sector we have the following PQ charges: 
\begin{eqnarray} 
\eta^0:\; & &-2X_u  = 2X_d =X_\nu-X_{\nu_R},\nonumber \\ 
\eta^-_1:\;  & &-(X_u+X_d)=X_u+X_d =X_l-X_{\nu_R},\nonumber \\ 
\eta^+_2:\; & &-(X_J+X_u) = X_j+X_d = -(X_l+X_{\nu_R}), 
\nonumber \\ 
\rho^0:\; & &2 X_u = -2 X_d  =X_l-X_{E_R},\nonumber \\ 
\rho^+: \;& &-(X_u+X_d) = X_u+X_d  
=X_\nu-X_{E_R} , 
\nonumber \\ 
\rho^{++}:\; & &- (X_J+X_d) = X_j+X_u = -(X_l+X_{E_R}), 
\nonumber \\ 
\chi^{\prime-}:\; & &-(X_u+X_J) =X_d+X_j =X_\nu+X_{E_L},\nonumber \\ 
\chi^{--}:\; & &-(X_d+X_J) =X_u+X_j = X_{E_L}+X_l,\nonumber \\  
\chi^0: \;& &-2X_J = 2X_j=X_{E_L}-X_l,\nonumber \\ 
 \phi: \;& &-2 X_j, \nonumber \\  
\sigma^0_1:\; & & X_d+3X_j=2X_\nu,\nonumber \\ 
h^-_1:\;  & & 3X_j-X_d=X_l+X_\nu,\nonumber \\ 
h^+_2:\; & & 4X_j = -X_l+X_\nu, 
\nonumber \\ 
\sigma^0_2:\; & & 4X_j -2X_d =0, \nonumber \\ 
H^{--}_1: \;& & 3(X_j-X_d)= 2X_l, 
\nonumber \\ 
H^{++}_2:\; & & 5X_j-X_d =-2X_l. 
\label{pq23} 
\end{eqnarray} 
From Eqs.~(\ref{pq23}) we obtain the relations 
$X_j=-X_J=\frac{1}{2}X_d=-\frac{1}{2}X_u=-\frac{2}{3}X_l= 
-\frac{2}{3}X_{\nu_R} 
=\frac{2}{5}X_\nu=\frac{2}{5}X_{E_R}=2X_{E_L}=-\frac{1}{2}X_\phi$. 
 
Notice that the mass scale related to  the exotic charged lepton $E$, up 
to an arbitrary dimensionless constant $G_E\sim O(1)$, is related 
to  the PQ energy scale since the requirement of the symmetries of the 
model imposes the Yukawa couplings in Eq.~(\ref{yu2}). 
 
Notice also that, at the energy scale below the breakdown of the $SU(3)_L$ symmetry,
this model has  scalar multiplets transforming under $SU(3)_C\otimes
SU(2)_L\otimes U(1)_Y$ as follows: two doublets
$(\rho^+,\,\rho^0)\sim({\bf1},{\bf2},+1)$,
$(\eta^0,\,\eta^-)\sim({\bf1},{\bf2},-1)$ and a  
non-hermitian triplet $(H^{--},\,h^-_1,\sigma^0_1)\sim({\bf1},{\bf3},-2)$.
With the lighter scalar multiplets, and the usual degrees of freedom, the
energy scale at which $\sin^2\theta_W=0.25$ is 5.2 TeV. 
The doublets $(\chi^-,\,\chi^{--})\sim({\bf1},{\bf2},-2)$,
$(h^+_2,h^0)\sim({\bf1},{\bf2},+1)$ and the extra vector bosons have masses
proportional to $v_\chi$; the lepton singlet $E$ has a mass of the order of
$v_\phi$. More details will be given elsewhere.

\subsection{Model with three scalar triplets (Model B)} 
\label{subsec:modelb} 
 
In this model we do not introduce the scalar sextet and the Yukawa interactions
are
\begin{eqnarray} 
-{\cal L}^l_Y&=&H^\nu_{ab}\overline{(\Psi)}_{aL} 
\nu_{bR}\,\eta +H^l_{ab}\epsilon_{ijk} 
\overline{(\Psi)^c}_{iaL}\Psi_{jbL}\eta_k 
\nonumber \\ &+&K_a\overline{\Psi}_{aL}E_R\rho + 
K^\prime_a \chi^T \;\overline{E_L}\,(\Psi_{aL})^c+G_E\bar{E}_LE_R\,\phi^* 
\nonumber \\ & +&H.c.,  
\label{yu3}  
\end{eqnarray} 
where $H^l_{ab}$ is now an antisymmetric matrix. 
In both Yukawa interactions above, a general mixing is allowed in each 
charge sector. As  in the previous model, neutrinos are strictly  
Dirac particles. The charged leptons gain mass as in Ref.~\cite{seesaw}.  
 
If we want to implement a given texture for the quark 
and lepton mass matrices we have to introduce more scalar triplets, and a larger 
$Z_N$ symmetry will be possible in the model.  
 
Let us introduce a $Z_4$ symmetry with parameters denoted by  
$\tilde{\omega}_0$, $\tilde{\omega}_1$, $\tilde{\omega}^{-1}_1$, and 
$\tilde{\omega}_2\equiv \tilde{\omega}^{-1}_2$. $u_{\alpha R}$, $Q_{iL}$, and 
$\nu_{aR}$ transform with $\tilde{\omega}_1$; $d_{\alpha R}$, $Q_{3L}$, 
$\Psi_{aL}$, $E_R$, $\chi$, and $\phi$ transform with $\tilde{\omega}^{-1}_1$, 
$\eta$ transform with $\tilde{\omega}_2$, and all the other fields remain
invariant, i.e., transform with $\tilde{\omega}_0$.   After $Z_4$ is
imposed, the total lepton number $L$ and the PQ and  $Z_{11}$ symmetries are  
all automatically implemented in the Yukawa sector and in the scalar potential. 
The most general scalar potential is then  
\begin{equation} 
V^{(B)}_{\rm 331} =  V_{\rm H} + 
\left(\lambda\,\phi\,\epsilon^{ijk}\eta_i\rho_j\chi_k +  
\mbox{H. c.}\right).  
\label{peb} 
\end{equation} 
 
The following $Z_{11}$ symmetry is automatically implemented in both the Yukawa 
interactions and in the scalar potential: 
$Q_{iL}\to \omega_3Q_{iL}$, $Q_{3L}\to\omega_0Q_{3L}$,  
$u_{\alpha R} \to \omega_4u_{\alpha R}$,  
$d_{\alpha R}\to \omega^{-1}_1 d_{\alpha R}$,  
$J_R\to \omega^{-1}_5 J_R$, $j_{mR}\to\omega^{-1}_3 j_{mR}$,  
$\Psi_L\to \omega_2\Psi_L$, $E_L\to\omega_3E_L$, 
$\nu_R\to\omega^{-1}_5\nu_R$, $E_R\to\omega_1E_R$, 
$\eta\to \omega^{-1}_4\eta$, $\rho\to \omega_1\rho$, 
$\chi\to \omega_5\chi$, $\phi\to\omega^{-1}_2\phi$. 
It happens that, in addition to the $Z_{11}$ symmetry, the $U(1)_{\rm PQ}$ and the 
conservation of the total lepton number are also 
automatic i.e., a consequence of the gauge symmetry and renormalizability
of the model, in the interactions in Eqs.~(\ref{yu1}), (\ref{yu3}), and
(\ref{peb}). The PQ charge assignments for the fermions in the model are
as in Eq.~(\ref{pq1}); and in the scalar sector we have constraints
equations as in the previous subsection. In this case proceeding as in
the model A, we obtain the relations $X_d=X_u=0$ and
$X_l=X_{\nu_R}=X_\nu=X_{E_R}=-\frac{1}{2}X_j= 
\frac{1}{2}X_J=-\frac{1}{3}X_{E_L}=\frac{1}{4}X_\phi$. 
Notice that for leptons the PQ transformations are not a chiral symmetry. 
As in Model A, we see from Eq.(\ref{yu3}) that the mass scale related to the 
singlet charged lepton $E$ is related to $v_\phi$. 
Moreover, in this model we have that $\sin^2\theta_W=0.25$ at 4 TeV.

\section{conclusions} 
We have built two invisible axion models in which the axion is 
naturally light and protected against quantum gravity effects. In model A, the 
$Z_{13}$ symmetry has to be imposed but in model B the $Z_{11}$ symmetry is 
automatically implemented in the classical Lagrangian after imposing a $Z_4$ 
symmetry. With a $Z_{13}$ symmetry the axion is protected from gravitational 
effects even if $v_\phi\approx10^{12}$ GeV but, with a $Z_{11}$ symmetry, 
$v_\phi\lesssim10^{10}$ GeV. In both models the PQ symmetry is automatically 
implemented in the classical Lagrangian in the sense that it is not imposed on 
the Lagrangian but is just a  consequence of the particle content of the 
model, its gauge invariance, renormalizability, and Lorentz invariance.   
 
We would like to stress the strong constraint put on model building by the 
approach proposed in 
Refs.~\cite{axionsm,axion331}. Once the symmetry $Z_N$ is 
used, automatic or imposed, there is no choice for new interactions. In this 
vein, in both models neutrinos are strictly Dirac particles, and for this reason 
both models will be ruled out if the neutrinos turn to be   
Majorana particles, say by observation of the neutrinoless double beta decay.  
 
In the PQ solution to the strong CP problem the quark 
contributions to $\bar\theta$ are such that   
$\bar\theta\to \bar\theta-2\alpha\sum_fX_f$, where $f$ denotes any quark.     
In both models we have $\bar\theta\to \bar\theta-2\alpha X_j$ and we 
have the domain wall problem related to $Z_2\subset U(1)_{\rm 
PQ}$ for $X_j=2$.  
 
Concerning the CDM, we would like to call attention to a new possible 
candidate: a light and 
stable scalar in nonsupersymmetric models~\cite{cc,ft}. Although the mass
of 
this scalar is in the range $32\,{\rm GeV}\,\lesssim m<M_Z$, the model is still 
compatible with the LEP data since the lightest scalar is almost a  
singlet under $SU(2)_L\otimes U(1)_Y$, and it may be mistaken for a light, 
$m_\chi\lesssim 50$ GeV~\cite{bottino}, or for  the usual, 
$m_\chi\stackrel{>}{\sim}50$ GeV, neutralino. We recall that the latter bound  
comes from LEP2 searches for the corresponding chargino 
$m_{\chi^\pm}\stackrel{>}{\sim} 100$ GeV.   
 
This work was supported by Funda\c{c}\~ao de Amparo \`a Pesquisa 
do Estado de S\~ao Paulo (FAPESP) and partially by Conselho Nacional de  
Ci\^encia e Tecnologia (CNPq).

\end{multicols}

\end{document}